\documentstyle[12pt,epsf,prd]{article}
\newcommand{\pbar}{\mbox{$\overline{p}$}}

\newcommand{\etal}{\mbox{\em et al.}}

\def\lessim{\mathrel {\vcenter {\baselineskip 0pt \kern 0pt
\hbox{$<$} \kern 0pt \hbox{$\sim$}}}}
\def\gepsfcentered#1{
  \def\testit{#1}
  \def\lbracket{[}
  \ifx\testit\lbracket
    \let\dofilecmd=\gepsfwithopt
  \else
    \let\dofilecmd=\gepsfnoopt
  \fi
  \dofilecmd}

\def\gepsfnoopt#1{
  \ifdofig
    \begin{center}
    \leavevmode
    \epsffile{#1}
    \end{center}
  \fi}

\def\gepsfwithopt#1 #2 #3 #4]#5{
  \ifdofig
    \addtocounter{figure}{1}
    \immediate\write16{Figure \thefigure(\thepage). #5 #1 #2 #3 #4}
    \write16{A Fig appears on page \thepage.}
    \addtocounter{figure}{-1}
    \begin{center}
    \leavevmode
    \gepsfmaxx=0.94\textwidth
    \epsffile[#1 #2 #3 #4]{#5}
    \end{center}
  \fi}

\newdimen\gepsfmaxx
\def\epsfsize#1#2{
  \ifnum \epsfxsize=0
    \ifnum \epsfysize=0
      \ifnum #1 > \gepsfmaxx
        \gepsfmaxx
      \else
        #1
      \fi
    \else
      \epsfxsize
    \fi
  \else
    \epsfxsize
  \fi
}

\begin{document}
\pagestyle{empty}
%
\begin{flushright}
FERMILAB-PUB-98/216-E\\
July 21, 1998
\end{flushright}
\footnotetext[1]{Submitted to Phys. Rev. Lett.}

\vspace{1.0cm}
\begin{center}
{\Large\bf Search for muonic decays of the antiproton at
the Fermilab Antiproton Accumulator
\footnotemark[1]}
\end{center}
\vspace{0.5cm}

\begin{center}
\renewcommand{\baselinestretch}{1}
M. Hu, G.R. Snow \\
{\it University of Nebraska, Lincoln, Nebraska 68588}\\
\vspace{0.2cm}
S. Geer, J. Marriner, M. Martens, R.E. Ray, J. Streets, W. Wester \\
{\it Fermi National Accelerator Laboratory, Batavia, Illinois 60510}\\
\vspace{0.2cm}
T. Armstrong \\
{\it Pennsylvania State University, University Park, Pennsylvania 16802}\\
\vspace{0.2cm}
C. Buchanan, B. Corbin, M. Lindgren, T. Muller$^{\dag}$ \\
{\it University of California at Los Angeles, Los Angeles, California 90024} \\
\vspace{0.2cm}
R. Gustafson \\
{\it University of Michigan, Ann Arbor, Michigan 48109}\\
\end{center}

\vspace{0.1in}
\begin{center}
(The APEX Collaboration)
\end{center}
\vspace{0.5cm}
\begin{abstract}
\nopagebreak
A search for antiproton decay 
has been made at the Fermilab Antiproton Accumulator. Limits are 
placed on six antiproton decay modes which contain a final-state muon. 
At the 90\% C.L. we find that 
$\tau_{\overline{p}}$/$B(\overline{p}\rightarrow\mu^-\gamma) 
> 5.0 \times 10^4$~yr, 
$\tau_{\overline{p}}$/$B(\overline{p}\rightarrow\mu^-\pi^0) 
> 4.8 \times 10^4$~yr, 
$\tau_{\overline{p}}$/$B(\overline{p}\rightarrow\mu^-\eta) 
> 7.9 \times 10^3$~yr, 
$\tau_{\overline{p}}$/$B(\overline{p}\rightarrow\mu^-\gamma\gamma) 
> 2.3 \times 10^4$~yr, 
$\tau_{\overline{p}}$/$B(\overline{p}\rightarrow\mu^-K^0_S) 
> 4.3 \times 10^3$~yr, and  
$\tau_{\overline{p}}$/$B(\overline{p}\rightarrow\mu^-K^0_L) 
> 6.5 \times 10^3$~yr.
\end{abstract}
\vspace{1.0cm}

\clearpage
\pagestyle{plain}
\setcounter{page}{2}

The CPT theorem requires that the proton and antiproton 
($\overline{p}$) lifetimes are equal. 
Searches for proton decay have yielded lower limits on 
the proton lifetime $\tau_p > O(10^{32})$~yr~\cite{pdb}. 
A search for $\overline{p}$ decay with a short lifetime 
($\tau_{\overline{p}} << \tau_p$) tests both the CPT 
theorem and the intrinsic stability of antimatter. 

Currently, the most stringent lower limits on $\tau_{\overline{p}}$ 
can, in principle, be obtained from a comparison of recent measurements 
of the cosmic ray antiproton flux with predictions based on expectations 
for secondary production of antiprotons in the interstellar medium. 
The agreement between the observed and predicted rates implies 
that $\tau_{\overline{p}}$ is not small compared to T/$\gamma$, where 
T is the $\overline{p}$ confinement time within the galaxy ($\sim 10^7$~yr) 
and $\gamma$ is the Lorentz factor for the observed antiprotons. 
However, to extract a limit on $\tau_{\overline{p}}$ requires 
careful consideration of the relationship between the interstellar 
$\overline{p}$ flux and the flux observed at the Earth, and an 
adequate estimate of the systematic uncertainties on the predictions. 
When completed these calculations are expected to yield 
a limit in the range $\tau_{\overline{p}} > 10^5$--$10^6$~yr~\cite{dallas}. 
This indirect limit would not be valid if current models of $\overline{p}$ 
production, propagation, and interaction in the interstellar medium are 
seriously flawed by, for example, our lack of knowledge of the nature of most 
of the matter in the galaxy.

Laboratory searches for $\overline{p}$ decay have so far provided less 
stringent limits on $\tau_{\overline{p}}$. However, these limits do not 
suffer from large model dependent uncertainties. 
The most stringent published laboratory limits on $\overline{p}$ decay 
have been obtained from (a) 
measurements of the containment lifetime of $\sim 1000$ antiprotons 
stored in an ion trap, yielding~\cite{gabrielse} 
$\tau_{\overline{p}} > 3.4$~months, and (b) prior 
searches at the Fermilab antiproton accumulator 
for explicit $\overline{p}$ 
decay modes with an electron in the final state, 
yielding~\cite{t861_prl}  
$\tau_{\overline{p}}/B(\overline{p} \rightarrow e^-\gamma) > 1848$~yr, 
$\tau_{\overline{p}}/B(\overline{p} \rightarrow e^-\pi^0) > 554$~yr,
$\tau_{\overline{p}}/B(\overline{p} \rightarrow e^-\eta) > 171$~yr,
$\tau_{\overline{p}}/B(\overline{p} \rightarrow e^-K^0_S) > 29$~yr, and 
$\tau_{\overline{p}}/B(\overline{p} \rightarrow e^-K^0_L) > 9$~yr. 

Angular momentum conservation requires that a decaying $\overline{p}$ would  
produce a fermion (electron, muon, or neutrino) in the final state. 
In this paper we report results from the first search for antiproton decay 
modes with a muon in the final state. The results have been obtained using 
the APEX detector at the Fermilab Antiproton Accumulator. The experiment 
was designed to search for decay modes with an electron or muon 
in the final state. The results from the search for 
$\overline{p} \rightarrow e^- + X$ will be described elsewhere~\cite{inprep}. 

\begin{figure}
\epsfxsize6.in
\centerline{\epsffile{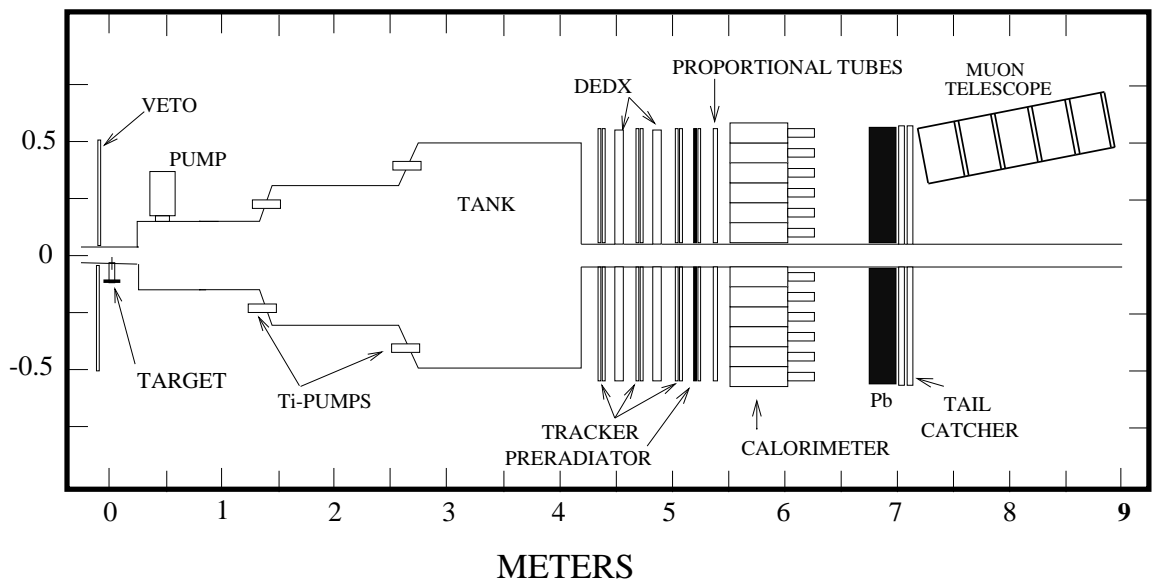}}
\vspace{0.5cm}
\caption{Schematic of the APEX detector.}
\label{detector_fig}
\end{figure}

A full description of the APEX detector can be found in Ref.~\cite{nim}. 
In the following we give a brief description of the main detector components 
relevant to the analysis described in this paper. 
A schematic of the APEX detector is shown in Fig.~\ref{detector_fig}. 
The detector, which was located in a straight section of the
Accumulator ring, consisted of:
(i) A 3.7~m long evacuated decay tank operated at $10^{-11}$~Torr. 
The downstream section of the tank consists of a 96~cm diameter cylinder 
with a 1.2~mm thick stainless steel window. 
(ii) A movable tungsten wire target at the upstream end of the tank 
($z$ = 0, 
where in the APEX co-ordinate system $z$ is measured in the direction of the
antiproton beam). 
The target could be inserted into the beam halo to produce a localized 
source of particles for aligning and calibrating the detector.
(iii) Horizontal and vertical scintillation counters arranged 
around the 10~cm diameter beam pipe immediately upstream of 
the tank and target. The counters covered a \mbox{$1 \times 1$~m$^2$} 
area normal to the 
beam direction, and were used to veto tracks from upstream interactions. 
(iv) Three planes of horizontal and three 
planes of vertical scintillation counters downstream of the tank. 
The last planes of horizontal and vertical counters 
were downstream of a 2.3 radiation length lead wall, providing a preradiator 
to aid in identifying electrons and photons. The remaining counter planes 
were upstream of the lead, and provided pulse height information used to 
count the number of charged particles in an event (dE/dx counters). 
(v) Three planes of horizontal and three planes of vertical 2~mm diameter 
scintillating fibers downstream of the tank and upstream of the preradiator 
lead. These detectors were used for tracking, and provided three space points 
along the track trajectory with typical residuals of 620~$\mu$m in the 
directions transverse to the beam direction. This enabled the origin of tracks 
emerging from the decay tank to be located with a precision
$\sigma_z \sim 13$~cm along the beam orbit. 
The measured average single-hit efficiency for the tracker planes is 
($89.2 \pm 1.5$)\% for tracks in the event samples described in this paper. 
The measured track reconstruction efficiency, based on a sample of 
events that are consistent with having one minimum-ionizing particle 
passing through the dE/dx and muon telescope (see below) counters, is 
($90 \pm 7$)\%. (vi) A lead--scintillator sampling electromagnetic 
calorimeter~\cite{ref:fcal} constructed from 144 rectangular 
$10 \times 10$~cm$^2$ modules that are 14.7 radiation lengths deep. 
The modules were arranged in a $13 \times 13$ array with 6 modules 
removed from each of the 
four corners, and the central module removed to allow passage of the 
beam pipe. The calorimeter was calibrated by measuring the response to 
minimum ionizing tracks and reconstructing $\pi^0 \rightarrow \gamma\gamma$ 
and $\eta \rightarrow \gamma\gamma$ mass peaks 
using data samples recorded with the calibration target inserted in the 
beam halo. 
The measured mass peaks had fractional r.m.s. widths given by
$\sigma_m$/m~$\sim 0.25$ for cluster pairs in the energy range of
interest for the experiment. This 
mass resolution is dominated by the energy resolution of the calorimeter. 
(vii) A tail catcher (TC) downstream of the calorimeter 
consisting of a 20~cm deep lead wall followed by planes of horizontal and
vertical scintillation counters. 
(viii) A limited-acceptance muon telescope (MT), 10
nuclear interaction lengths deep, located downstream of the TC, 
and aligned to point towards the center of the decay tank. 
The MT consists of a sandwich of five iron plates and five 
$30 \times 30$~cm$^2$ scintillation counters. 

The APEX experiment took data when there were typically $10^{12}$ antiprotons 
circulating in the 474~m circumference accumulator ring operating with a 
central $\overline{p}$ momentum of $8.90 \pm 0.01$~GeV/c 
($\gamma = 9.54 \pm 0.01$). 
A measure of the sensitivity of the APEX data sample is given by:
$$\displaystyle{S \;\equiv\; \frac{1}{\gamma} \int{N_{\pbar}\,(t)\,dt}
\;=\; (3.31 \pm 0.03) \times 10^{9}\,yr,}$$
where N$_{\pbar}\,(t)$ is the number of circulating antiprotons at time
$t$, the integral is over the live-time of the
experiment, and the uncertainty arises from the precision with which the 
time dependent beam current was recorded. 

To search for muonic $\overline{p}$ decays occurring as the beam particles 
traverse the decay tank, data were recorded with 
a muon trigger that required a coincidence between at least two of the five 
MT scintillation counters. These triggers were eliminated if 
they were in coincidence 
with a signal in one or more of the upstream veto counters indicating 
the presence of an  interaction upstream of the decay tank. 
This loose trigger resulted in 1.2 $\times 10^{6}$ events being recorded. 
These events predominantly arise from interactions of the
$\overline{p}$ beam with the residual gas in the decay tank or with
material surrounding the beam. The coincident MT counter signals are
then produced by traversing muons coming from 
charged pion decays, and by hadronic showers not
contained in the calorimeter and TC. 

Offline, after final calibration of the scintillation counters, 
the upstream veto counter requirement 
was re-applied using a more stringent threshold. 
This reduced the data sample to 1.1 $\times 10^{6}$ events. 
Further requirements were then imposed to select events containing 
a candidate energetic muon that traverses the MT and 
originates from within the decay tank. 
These requirements were (i) 
that at least four of the five MT counters be above threshold 
($4.2 \times 10^4$ events), and (ii) the presence of 
one and only one scintillating fiber track 
that extrapolates to the MT counters within the expected uncertainty
due to multiple scattering, and also 
extrapolates back to the beam orbit with a point of closest approach 
within the fiducial volume of the decay tank ($0 < z < 375$~cm) 
and with an impact parameter less than 1~cm (416 events). 

\begin{figure}
\epsfxsize6.in
\centerline{\epsffile{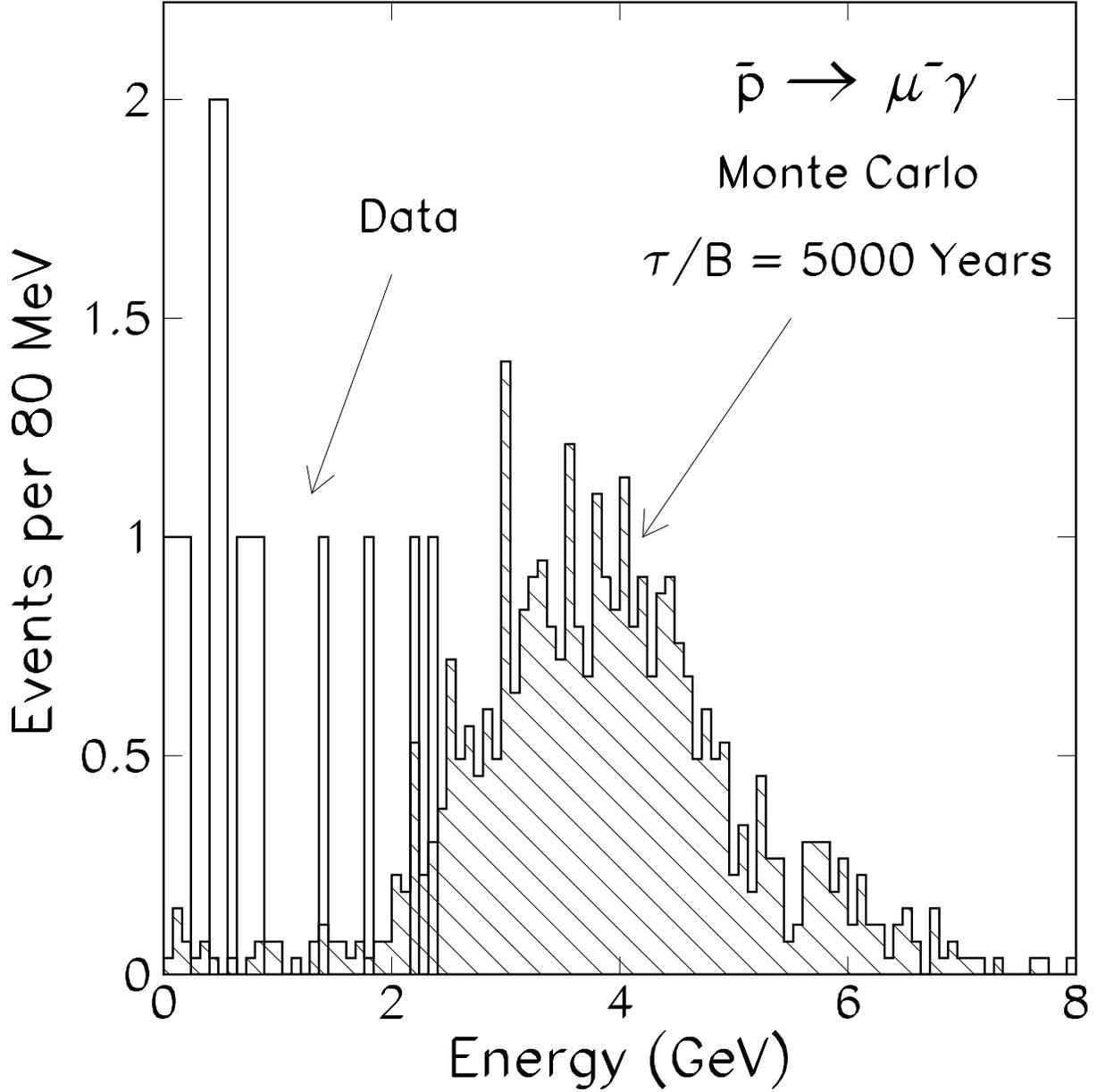}}
\vspace{-1.5cm}
\caption{Distribution of neutral cluster energies for the 
14 events that pass the $\overline{p} \rightarrow \mu^- \gamma$ 
selection criteria described in the text (open histogram) compared 
with the predicted distribution for a signal corresponding to 
$\tau_{\overline{p}}/B = 5000$~yr (hatched histogram). Limits are 
based on the region E $>$ 3 GeV.
}
\label{mugam_fig}
\end{figure}

To further suppress backgrounds additional requirements can be imposed 
on the event topology and kinematics. 
We begin by considering the process 
$\overline{p} \rightarrow \mu^{-}\gamma$, which 
would result in events in which an 
energetic photon is produced coplanar with the muon (i.e. traveling 
within the plane defined by the muon and the incoming beam direction). 
After requiring at least one neutral cluster~\cite{nim} in the 
calorimeter (209 events) 
that is coplanar with the muon candidate ($\pm 5^\circ$), the data sample 
is reduced to 14 events. 
The observed neutral cluster energy distribution for these 
events is compared in Fig.~\ref{mugam_fig} with the predicted distribution 
obtained using the 
GEANT~\cite{geant} simulation described below, and corresponding to 
$\overline{p} \rightarrow \mu^-\gamma$ decay with a lifetime 
$\tau_{\overline{p}}$/B$(\overline{p} \rightarrow \mu^-\gamma) = 5000$~yr. 
The observed distribution peaks at low cluster energies, with a tail 
extending to approximately 2~GeV. In contrast, the predicted distribution 
for $\overline{p} \rightarrow \mu^-\gamma$ decays peaks at about 3.5~GeV, 
with only 4.4\% of the simulated events having cluster energies less than 
2~GeV. We conclude that there is no evidence for a signal. To eliminate 
the background we therefore require that the neutral cluster energy exceed 
a minimum value $E_{min}$, and choose $E_{min} = 3$~GeV. No candidate events 
remain. We note that although the choice of $E_{min}$ is somewhat arbitrary, 
the final limit that we obtain on the decay 
$\overline{p} \rightarrow \mu^-\gamma$ is not very sensitive 
to small changes in 
$E_{min}$.
The resulting limit on 
$\tau_{\overline{p}}$/B($\overline{p} \rightarrow \mu^-\gamma$) 
is given in years by 
\begin{equation}
\tau_{\overline{p}}/B(\overline{p} \rightarrow \mu^-\gamma) \; > \;  
\frac{\epsilon}{\gamma} \; \frac{1}{N_{max}} \; \int{N_{\pbar}\,(t)\,dt} 
\; = \; (3.31 \pm 0.03) \times 10^9 \; \frac{\epsilon}{N_{max}}\; ,
\end{equation}
where $N_{max}$ = 2.3 is the 90\% C.L. upper limit
on the observation of $N = 0$ events, and 
$\epsilon$ is the fraction of decays taking place uniformly 
around the accumulator ring that would pass the trigger and event 
selection requirements. To take account of $\sigma_{\epsilon}$, 
the systematic uncertainty on 
$\epsilon$, we use the prescription 
given in Ref.~\cite{cousins}, giving at 90\% C.L. 
\begin{equation}
N _{max} \; = \; 2.3 \times (1 + 2.3~\sigma_r^2/2) \; ,
\end{equation}
where $\sigma_r \equiv \sigma_\epsilon/\epsilon$. 

The GEANT Monte Carlo program has been used to simulate the detector 
response and calculate $\epsilon$. The detector simulation includes a 
full description of the detector geometry, and correctly 
describes the calorimeter, tracker, and relevant scintillation counter 
responses (dE/dx, MT) measured using calibration data, together 
with the measured performance of the muon trigger. Further details can 
be found in Ref.~\cite{nim}. 
The efficiency $\epsilon$ was obtained by 
generating $10^5 \; \overline{p} \rightarrow \mu^{-}\gamma$ decays 
uniformly along 
the beam orbit within the decay tank. To a good approximation the geometrical 
efficiency of the detector and trigger is negligible for decays occurring 
outside of the tank. We obtain $\epsilon = (3.7 \pm 0.9) \times 10^{-5}$, 
where the uncertainty on $\epsilon$ arises from the systematic 
uncertainties on the trigger calibration, calorimeter energy scale, 
and track efficiency. The trigger and calorimeter scale uncertainties 
yield contributions to $\sigma_{\epsilon}$ of $\pm18$\% and $\pm 
15$\%, respectively, and have been evaluated by analyzing GEANT 
Monte Carlo $\overline{p} \rightarrow \mu^{-}\gamma$ 
samples in which the simulated trigger and 
calorimeter scales have been changed by $\pm1\,\sigma$. The overall 
systematic uncertainty on $\epsilon$ has been calculated by adding 
these contributions in quadrature with the uncertainty on the track 
efficiency ($\pm7$\%). 
Inserting the calculated $\epsilon$ and $\sigma_\epsilon$ into Eqs.~1 and 
2 we obtain the result 
$\tau_{\overline{p}}/B(\overline{p} \rightarrow \mu^-\gamma) \; > \;$ 
$5.0 \times 10^4$~yr (90\% C.L.). 

Now consider the two--body decays $\overline{p} \rightarrow \mu^-\pi^0$ and  
$\overline{p} \rightarrow \mu^-\eta$ (with 
$\pi^0 ,\; \eta \rightarrow \gamma\gamma$), and the three--body decay 
$\overline{p} \rightarrow \mu^-\gamma\gamma$. 
These decays would result in 
events with one or two neutral clusters observed in the calorimeter, where 
the one--cluster events occur when the two photon--showers are not spatially 
resolved in the calorimeter or when one of the photons is outside of the 
calorimeter acceptance. In addition, the $\eta$ may also decay into more 
complicated final states producing further clusters in the calorimeter. 
To optimize the search for $\mu^-\pi^0$ and $\mu^-\eta$ final states we 
divide the 209 events described previously that have a muon 
candidate plus one or more neutral clusters into two 
subsamples, namely a one--cluster sample containing 104 events, and 
a multi--cluster sample containing 105 events. 
The clusters in the one--cluster sample are required to be coplanar with the 
muon ($\pm 10^\circ$), and the cluster--pairs formed by the two highest energy 
clusters in the multi--cluster events are also required to be coplanar with the 
muon ($\pm 10^\circ$). These requirements reduce the samples to 
13 one-cluster and 10 multi-cluster events. We next require that the 
multi-cluster events, which would contain 2 photons if they were 
genuine $\overline{p} \rightarrow \mu^-\pi^0, \mu^-\eta$, 
or $\mu^-\gamma\gamma$ decays, have 
a preradiator signal above a threshold of 0.5 $\times$ 
minimum-ionizing pulse height. Only 3 multi-cluster events satisfy 
this requirement. Hence we are left with 16 events with $\ge$~1 cluster
events ($\mu \;+ \ge 1$~cluster) for further analysis. 

\begin{figure}
\epsfxsize6.in
\centerline{\epsffile{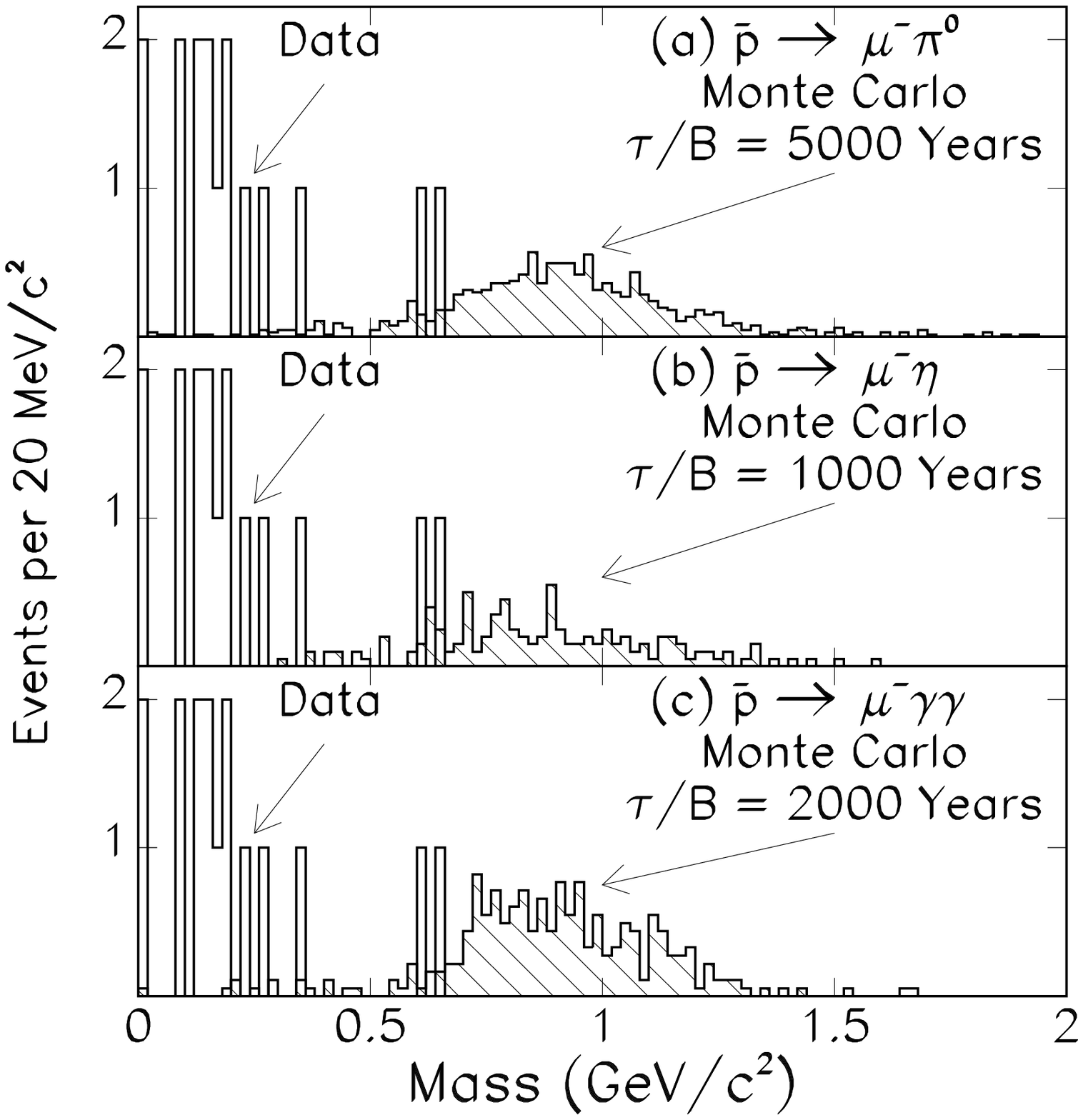}}
\vspace{-1.5cm}
\caption{Distribution of event masses for 
the 16 events that pass the selection criteria described in the text 
(open histograms) compared with predictions from a GEANT simulation 
(hatched histograms) for the decays 
(a) $\overline{p} \rightarrow \mu^-\pi^0$, 
(b) $\overline{p} \rightarrow \mu^-\eta$, and 
(c) $\overline{p} \rightarrow \mu^-\gamma\gamma$. The two entries in the 
lowest bin are events with calorimeter energy less than the $\pi^0$ 
rest energy. The predicted signal distributions are normalized to 
correspond to $\tau_{\overline{p}}/B = 5000$~yr for the $\mu^-\pi^0$ mode, 
1000 yr for the $\mu^-\eta$ mode, and 2000~yr for the $\mu^-\gamma\gamma$ mode. 
Limits are based on the region Mass $>$ 0.75 GeV.
}
\label{mass_fig}
\end{figure}

Under the hypothesis that the observed ($\mu\gamma$) and 
($\mu\gamma\gamma$) systems arise from the decay of a beam particle, the 
mass of the parent particle can be computed from the measured muon 
direction and 
the directions and energies of the neutral clusters, using the constraint 
that the vector sum of the momentum components of the daughter particles 
transverse to the beam direction is zero. The resulting mass distribution 
for the remaining 16 ($\mu \;+ \ge 1$~cluster) events is 
compared in Fig.~\ref{mass_fig} 
with predictions from the GEANT simulation for the decays 
(a) $\overline{p} \rightarrow \mu^-\pi^0$, 
(b) $\overline{p} \rightarrow \mu^-\eta$, 
and (c) $\overline{p} \rightarrow \mu^-\gamma\gamma$. 
The observed mass distribution peaks at low masses with a tail 
extending to approximately 0.7~GeV/c$^2$. In contrast to this, 
the simulated signal distributions peak at the $\overline{p}$ mass, 
with only 17\% (23\%) [13\%] of the simulated $\mu^-\pi^0$ 
($\mu^-\eta$) [$\mu^-\gamma\gamma$] decays resulting in reconstructed 
masses less than 0.7~GeV/c$^2$. 
We conclude that there is no evidence for a signal. To eliminate
the background we therefore require that the reconstructed mass exceed
a minimum value $m_{min}$, and choose $m_{min} = 0.75$~GeV/c$^2$. 
No candidate events
remain. We note that although the choice of $m_{min}$ is somewhat arbitrary,
the final limits that we obtain on the decays 
$\overline{p} \rightarrow \mu^-\pi^0$, $\mu^-\eta$, and $\mu^-\gamma\gamma$ 
are not very sensitive to small changes in $m_{min}$.
The calculated overall efficiencies 
$\epsilon$ for these decays to pass our trigger and analysis requirements 
are $(3.6 \pm 0.9) \times 10^{-5}$ for the $\mu^-\pi^0$ mode, 
$(6.1 \pm 1.9) \times 10^{-6}$ for the $\mu^-\eta$ mode, and 
$(1.8 \pm 0.5) \times 10^{-5}$ for the $\mu^-\gamma\gamma$ mode.
Substituting the values into Eqs.~1 and 2 yields the limits  
$\tau_{\overline{p}}/B(\overline{p} \rightarrow \mu^-\pi^0) \; > \;$
$4.8 \times 10^4$~yr, 
$\tau_{\overline{p}}/B(\overline{p} \rightarrow \mu^-\eta) \; > \;$
$7.9 \times 10^3$~yr at 90\% C.L., and 
$\tau_{\overline{p}}/B(\overline{p} \rightarrow \mu^-\gamma\gamma) \; > \;$
$2.3 \times 10^4$~yr at 90\% C.L.

Now consider the other possible two--body muonic decay modes of the
antiproton, namely decays into the final states $\mu^-K^0_S$, $\mu^-K^0_L$, 
$\mu^-\rho^0$, and $\mu^-\omega$. Detailed GEANT simulations have been made 
for these decay modes. 
The muons from the $\mu^-\rho^0$ and $\mu^-\omega$ decay modes are predicted 
to have energies that are too low to penetrate the muon telescope.
We therefore restrict 
ourselves to the $\mu^-K^0_S$ and $\mu^-K^0_L$ modes. 
The calculated efficiencies $\epsilon$ for these decays to satisfy the 
trigger and either the $\mu^-\gamma$ or $\mu^-\pi^0$ search criteria described 
previously are 
$(3.3 \pm 1.0) \times 10^{-6}$ for the $\mu^-K^0_S$ mode, and
$(5.0 \pm 1.5) \times 10^{-6}$ for the $\mu^-K^0_L$ mode.
Substituting these values into Eqs.~1 and 2 yields the limits 
$\tau_{\overline{p}}/B(\overline{p} \rightarrow \mu^-K^0_S) \; > \;$
$4.3 \times 10^3$~yr and
$\tau_{\overline{p}}/B(\overline{p} \rightarrow \mu^-K^0_L) \; > \;$
$6.5 \times 10^3$~yr at 90\% C.L. 

Finally, our results for the five muonic decay modes presented in this paper 
are summarized in Table~1, and a more comprehensive description of the 
analysis can be found in Ref.~\cite{huthesis}. 

\vspace{0.2cm}
The APEX experiment was performed at the Fermi National Accelerator 
Laboratory, which is operated by Universities Research Association,
under contract DE-AC02-76CH03000 with the U.S. Department of
Energy.
%

\clearpage

\begin{table}
\renewcommand{\baselinestretch}{1}
\label{tab:com}
\centering{
\caption{Summary of lifetime limits for $\overline{p}$ decay to six 
muonic final states with calculated efficiencies $\epsilon$ for each 
mode.}
\vspace{0.6 cm}
\begin{tabular}{lcc}  
\hline \hline
Decay Mode & $\epsilon$ & $\tau$/B Limit (yr)\\
           &            & (90 \% C.L.)\\
\hline
$\mu^- + \gamma$ & $(3.7 \pm 0.9) \times 10^{-5}$ & $> 5.0 \times 10^4$ \\
$\mu^- + \pi^0$  & $(3.6 \pm 0.9) \times 10^{-5}$ & $> 4.8 \times 10^4$ \\
$\mu^- + \eta$   & $(6.1 \pm 1.9) \times 10^{-6}$ & $> 7.9 \times 10^3$ \\
$\mu^- + \gamma\gamma$& $(1.8 \pm 0.5) \times 10^{-5}$ & $> 2.3 \times 10^4$ \\
$\mu^- + K^0_S$  & $(3.3 \pm 1.0) \times 10^{-6}$ & $> 4.3 \times 10^3$ \\
$\mu^- + K^0_L$  & $(5.0 \pm 1.5) \times 10^{-6}$ & $> 6.5 \times 10^3$ \\
\hline \hline
\end{tabular}
}
\end{table}

\clearpage

%
%

\vspace{0.5cm}
\begin{thebibliography}{99}
%
\bibitem[\dag]{muller} Present address: Universit\"{a}t Karlsruhe,
76128 Karlsruhe, Germany.

\bibitem{pdb} R.M. Barnett \etal, Physical Review {\bf D54}, 1 (1996)
     and 1997 off-year partial update for the 1998 edition available on
     the PDG WWW pages (URL: http://pdg.lbl.gov/).

\bibitem{dallas} D. Kennedy, private communication.

\bibitem{gabrielse} G. Gabrielse \etal, Phys. Rev. Lett., 
{\bf 65}, 1317 (1990).

\bibitem{t861_prl} S. Geer \etal, Phys. Rev. Lett., 
{\bf 72}, 1596 (1994).

\bibitem{inprep} APEX Collab., in preparation.

\bibitem{nim} T. Armstrong \etal (APEX Collab.), 
FERMILAB-PUB-97/379-E, accepted for publication 
in Nucl. Instrum. Methods A, December, 1997.

\bibitem{ref:fcal} M. A. Hasan \etal, Nucl. Instrum. Methods Phys. Res.,
{\bf A295}, 73 (1990).

\bibitem{geant} GEANT Version 3.21, R. Brun et al., CERN Program
Library.

\bibitem{cousins} R. Cousins and V. Highland, Nucl. Instrum. Methods 
Phys. Res., {\bf A320}, 331 (1992).

\bibitem{huthesis} M. Hu, Ph.D. Thesis, University of Nebraska, 1998.

\end{thebibliography}
\end{document}